\documentclass[aps,prb,floatfix,twocolumn,preprintnumbers,amsmath,amssymb,groupedaddress,showpacs,showkeys,superscriptaddress]{revtex4-1}

\usepackage{graphicx}  % Include figure files
\usepackage{bm}        % bold math
\usepackage{color}     % coloring
\usepackage{amsmath,amssymb}
\usepackage{nicefrac}
\usepackage{wasysym}
\usepackage{braket}  % braket brackets
\begin{document}

\title{Spin relaxation in fluorinated single and bilayer graphene}

\author{\surname{Susanne} Wellnhofer}
\affiliation{Institute for Theoretical Physics, University of Regensburg, 93040 Regensburg, Germany\\
 }
\author{\surname{Adam} Stabile}
\affiliation{Department of Physics, The Pennsylvania State University, University Park, PA 16802\\
 }
\author{\surname{Denis} Kochan}
\affiliation{Institute for Theoretical Physics, University of Regensburg, 93040 Regensburg, Germany\\
 }
\author{\surname{Martin} Gmitra}
\affiliation{Institute of Physics, P. J. \v{S}af\'{a}rik University in Ko\v{s}ice, 04001 Ko\v{s}ice, Slovakia\\
 }
\author{\surname{Ya-Wen} Chuang}
\affiliation{Department of Physics, The Pennsylvania State University, University Park, PA 16802\\
 }
\author{\surname{Jun} Zhu}
 \email{jxz26@psu.edu}
\affiliation{Department of Physics, The Pennsylvania State University, University Park, PA 16802\\
 }
\author{\surname{Jaroslav} Fabian}
\email{jaroslav.fabian@ur.de}
\affiliation{Institute for Theoretical Physics, University of Regensburg, 93040 Regensburg, Germany\\
}

\begin{abstract}
We present a joint experiment-theory study on the role of fluorine adatoms in spin and momentum scattering of charge carriers in dilute fluorinated graphene and bilayer graphene. The experimental spin-flip and momentum scattering rates and their dependence on the density of fluorine and carrier doping are obtained through weak localization and conductivity measurements, respectively, and suggest the role of fluorine as resonant magnetic impurities. For the estimated fluorine concentration of a few 100 ppm, the observed  spin lifetimes are in the range of 1-10\,ps. Theoretically, we established tight-binding electronic structures of fluorinated graphene and bilayer graphene by fitting to density functional
  supercell calculations and performed a comprehensive analysis of the spin-flip and momentum scattering rates within the same devices, aiming to develop a consistent description of both scattering channels. We find that resonant scattering in graphene is very sensitive to the precise position of the resonance level, as well as to the magnitude of the exchange coupling between itinerant carriers and localized spins. The experimental data point to the presence of weak spin-flip scatterers that, at the same time, relax the electron momentum strongly, nearly preserving the electron-hole symmetry. Such scatterers would exhibit resonance energies much closer to the neutrality point than what density functional theory predicts in the dilute limit. The inclusion of a magnetic moment on fluorine adatoms allowed us to qualitatively capture the carrier density dependence of the experimental rates but predicts a greater (weaker) spin (momentum) relaxation rate than the measurements. We discuss possible scenarios that may be responsible for the discrepancies. Our systematic study exposes the complexities involved in accurately capturing the behavior of adatoms on graphene.
\end{abstract}

\date{\today}
\maketitle

%-------------------------------------------------------------------------------------------
\section{Introduction}
%-------------------------------------------------------------------------------------------

Surface functionalization, which exploits the “all-surface” nature of two-dimensional atomically thin layers, is a powerful tool to engineer desired properties absent in pristine materials. Adatoms and molecular groups on graphene, for example, are shown to induce a band gap, modify its optical emission and enhance its solubility in aqueous solution~\cite{Sofo2007,Boukhvalov2009,Craciun2013:JPCM}.

Chemisorbed adatoms, such as H, introduce isolated magnetic moments \cite{Yazyev:RepProgPhys2010,Gonzalez-Herrero2016} and strong local spin-orbit coupling (SOC) to graphene~\cite{Min:PRB2006,Neto:PRL2009,Zhou:Carbon2010,Qiao:PRB2010,Weeks:PRX2011,Gmitra:PRL2013,Irmer2015:PRB,Zollner:Meth2016,Frank:PRB2017}. Owing to the gapless Dirac bands, adatoms on graphene and bilayer graphene can form sharp impurity states situated close to the charge neutrality point. As a result, the interaction between the impurities and the mobile carriers is resonantly enhanced~\cite{Stauber2007,Ferreira:PRB2011,Monteverde2010,Robinson2008}. The impurity's resonant nature depends on the valence orbitals and the adsorption site of the adatom~\cite{Ruiz2016,Uchoa2014,Weeks:PRX2011,Duffy2016,Irmer2018:PRB}. We adopt the
terminology that {\it resonant impurities} are to be distinguished from {\it strong midgap scatterers} which are described by a deep potential well of finite radius~\cite{Stauber2007,Chen2009,Ferreira:PRB2011} and therefore influence the charge scattering sector much more than simple vacancies~\cite{Irmer2018:PRB}. Both vacancies and strong midgap scatterers induce resonance levels directly at the charge neutrality point.

Due to the aforementioned resonant enhancement, adatom-induced magnetic moments on graphene can be a very effective source of spin-flip scattering, the unintentional presence of which provides a possible explanation for ultrafast spin relaxation in pristine graphene devices~\cite{Kochan2014:PRL,Kochan2015:PRL}. Local SOC induced by adatoms can also be a source for spin relaxation and manipulation~\cite{Bundesmann:PRB2015}. Engineering adatoms thus provides a potential route to instill magnetic and spintronic functionalities in graphene. Here, fluorination provides an attractive opportunity. For example, first-principles calculations predict a sizable local SOC of about 10 meV in dilute fluorinated graphene\cite{Irmer2015:PRB}.

Experimentally, the fluorination of graphene is relatively straightforward. Heavily fluorinated graphene exhibits a large band gap~\cite{Cheng2010,Nair2010,Robinson2010,Wang2014} and is spin-half paramagnetic~\cite{Nair2012:NP}. In the dilute limit, fluorination on single layer and bilayer graphene (SLG and BLG) induces strong midgap state scattering in the measured conductivity~\cite{Hong2012:PRL,Hong2011:PRB,Stabile2015}. Using weak localization (WL) as a probe, previous experiments by some of us also uncovered an anomalous large dephasing rate $\tau_{\phi}^{-1}$ in fluorinated SLG~\cite{Hong2012:PRL}. This observation points to the existence of fluorine-induced magnetic moments, similar to hydrogenated graphene~\cite{McCreary:PRL2012}, although a quantitative and mechanistic assessment has yet to be made. Unlike hydrogen~\cite{Yazyev:RepProgPhys2010,Sofo:PRB2012}, the formation of a magnetic moment in fluorinated graphene remains inconclusive among first-principle studies~\cite{Kim2013:PRB,Santos2012:NJP} presumably due to the self-interaction error in the exchange-correlation functionals~\cite{Mori2008,Casolo2010}. Furthermore, both fluorine concentration and carrier doping~\cite{Yndurain:PRB2014} can play a role, making the magnetic properties of fluorine adatoms a complex issue to address.

In this joint experiment-theory study we attempted to provide a \emph{quantitative} model to simultaneously capture the effect of fluorine adatoms on both spin and charge scattering. We are motivated by the availability of a complete set of conductivity and WL measurements on fluorinated SLG and BLG devices. These come from our previous studies~\cite{Hong2012:PRL,Hong2011:PRB,Stabile2015} and new data reported in Sec.~\ref{sec:exp}.  Theoretical investigations are built upon our previous calculations discussing the spin~\cite{Kochan2014:PRL,Kochan2015:PRL} and momentum~\cite{Irmer2018:PRB} relaxations of resonant impurities in graphene. Fluorinated SLG and BLG are described in a tight-binding (TB) model motivated by density functional theory (DFT) calculations.

Our systematic comparison of measurement and modeling revealed several insights. The theory and experiment  consistently describe the carrier density dependence of the scattering rates, supporting the resonant scattering mechanism for both spin and momentum relaxation. But we also find considerable quantitative differences, with our theory finding a greater spin relaxation rate than the experiment and underestimating the momentum relaxation rate. The high experimental momentum relaxation rate is consistent with fluorine being a strong midgap scatterer with electron-hole symmetry while our DFT results on large supercells of fluorinated graphene show a broad resonance away from the charge neutrality point. In plain terms, the DFT induced
exchange coupling is much greater, while the position of the resonance level far off, 
than what would be needed to account for the measured data. The difficulty to capture the experimental observations has motivated us to examine several potentially relevant scenarios. In particular, the varying local curvature of the graphene sheet may play an important role, the effect of which on the electronic structure and magnetic screening of fluorine should be carefully examined. Similarly, we see a need to investigate the magnetic and momentum scattering of adatom clusters, as they could produce resonances close to the charge neutrality point and thus a better match to the strong midgap scatterer model than individual adatoms. In addition, it is worthwhile to reexamine whether it is appropriate to use an independent scattering approximation for weakly resonant states such as fluorine in our DFT calculations. We hope that our work stimulates further studies in these directions.

The paper is organized as follows. Section~\ref{sec:exp} describes the new WL data in fluorinated BLG, while Sec.~\ref{sec:model} introduces the DFT and TB band structures of fluorinated SLG and BLG and our methods to calculate the spin and momentum relaxation rates. Results, comparison to experiments and discussion are presented in Sec.~\ref{sec:modExp}. Here, we point out the major differences between model and experiment and speculate on possible reasons. We conclude in Sec.~\ref{sec:conclude}.

%-------------------------------------------------------------------------------------------
\section{\label{sec:exp}Experiment}
The recipe used for fluorination and device fabrication and the characteristics of fluorinated SLG and BLG devices were described in previous studies \cite{Hong2011:PRB,Hong2012:PRL,Stabile2015}. References~\onlinecite{Hong2012:PRL} and \onlinecite{Stabile2015} found that a dilute fluorine adatom concentration gives rise to a momentum relaxation of charge carriers consistent with strong midgap scattering~\cite{Stauber2007,Ferreira:PRB2011,Stabile2015} which is characterized by resonance levels at the zero energy, i.e. the charge neutrality point, and consequently electron-hole symmetric conductivity $\sigma$, in agreement with experimental observations.

Here, we first describe new data on the density-dependent dephasing rate $\tau_{\phi}^{-1}(n)$ in fluorinated BLG. Measurements of $\sigma(n)$ and $\tau_{\phi}^{-1}(n)$ in the same devices enable us to investigate the effect of a single fluorine adatom on both charge and spin relaxation quantitatively in a self-consistent manner and in both SLG and BLG. This is the central objective of this work.

Fluorine concentrations of $n_\mathrm{F} = 2.2$, $3.8$, and $4.4 \times 10^{12}\,\mathrm{cm}^{-2}$ were obtained for the BLG devices W38, W02, and W03, respectively, using Raman spectroscopy and conductivity measurements in Ref.~\onlinecite{Stabile2015}. We obtain $\tau_{\phi}^{-1}(n)$ using magneto-conductance measurements $\sigma_s(B)$ similar to that described in Ref.~\onlinecite{Hong2012:PRL} in the carrier density regime of $n > n_\mathrm{F}$ in each device. The WL expression for BLG \cite{Gorbachev:PRL2007} accurately describes our magneto-conductance data, from which we determine the phase decoherence length $l_\phi$ and subsequently the dephasing rate $\tau_{\phi}^{-1}$ (see Appendix~\ref{app:exp}). We obtained $\tau_{\phi}^{-1}(n)$ over a range of carrier densities in the $10^{12}$-$10^{13}\,\mathrm{cm}^{-2}$ regime at a fixed temperature $T=1.7\,\mathrm{K}$.  We have also obtained through extrapolation the $T=0$ limit $\tau_{\rm sat}^{-1}(n)$ in W03 by a temperature dependence study (see Figure~\ref{fig:expSheetCondB} in Appendix~\ref{app:exp}).

\begin{figure}
  \includegraphics[width=0.8\columnwidth]{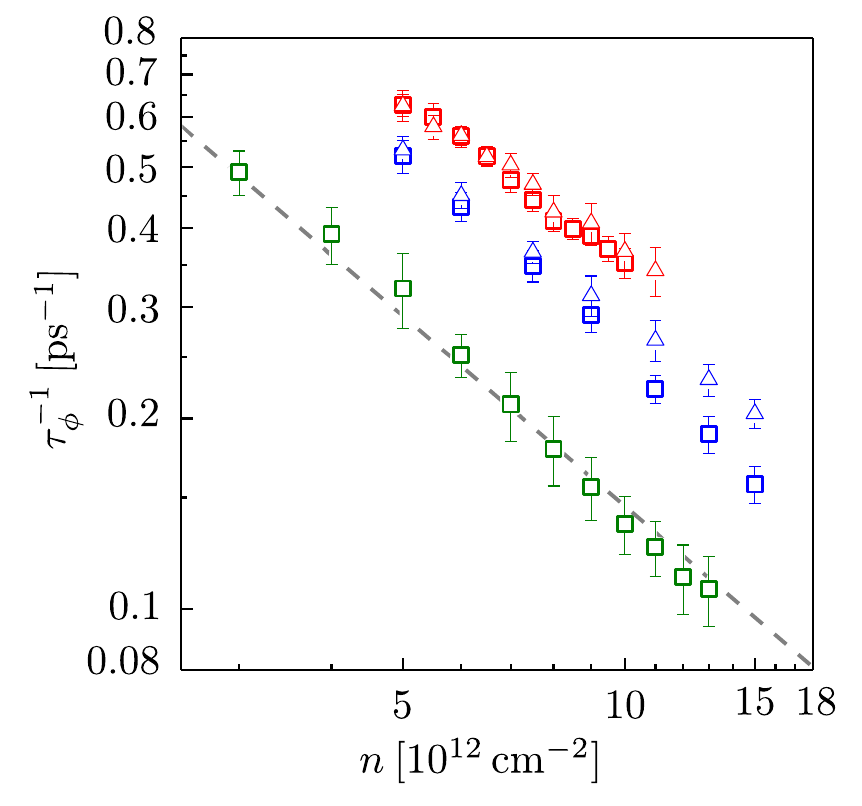}
  \caption{\label{fig:expTauPhi}Density-dependent dephasing rates vs. carrier density $\tau_{\phi}^{-1}(n)$ on a double-log plot in fuorinated BLG devices W03 (red, $n_{\rm F}=4.4\times 10^{12}\,\mathrm{cm}^{-2}$), W02 (blue, $n_{\rm F}=3.8\times 10^{12}\,\mathrm{cm}^{-2}$), and W38 (green, $n_{\rm F}=2.2\times 10^{12}\,\mathrm{cm}^{-2}$) at $T=1.7\,$K. Square symbols are for electrons and triangles are for holes. The gray dashed line corresponds to a power law dependence of $n^{-1}$.}
\end{figure}

Figure~\ref{fig:expTauPhi} plots $\tau_{\phi}^{-1}(n)$ of all BLG devices at $T=1.7\,$K. The magnitude of $\tau_{\phi}^{-1}$ ranges from $0.1$ to $1\,\mathrm{ps}^{-1}$ which is more than one order of magnitude larger than what is reported in the literature for pristine BLG \cite{Gorbachev:PRL2007}. Figure~\ref{fig:expTauPhi} also shows that  $\tau_{\phi}^{-1}$ is approximately electron-hole symmetric, as is the conductivity $\sigma(n)$ itself~\cite{Stabile2015}. Furthermore, $\tau_{\phi}^{-1}$ scales well with $n_{\rm F}$ and is well described by an empirical power law of $n^{-1}$. Following our earlier studies on fluorinated SLG, we tentatively attribute the enhanced $\tau_{\phi}^{-1}$ to spin-flip scatterings caused by fluorine-induced magnetic moments.

\begin{figure}
  \includegraphics[width=0.8\columnwidth]{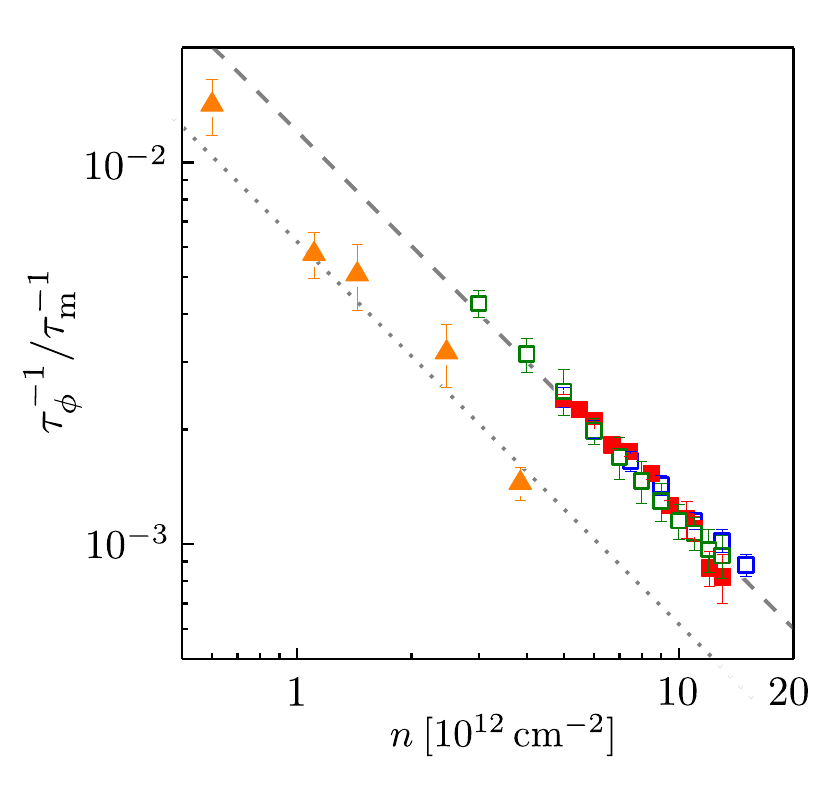}
  \caption{\label{fig:expTauPhiOverTauM}Experimental scattering rate ratio $\tau_{\phi}^{-1}/\tau_{\rm m}^{-1}$ versus carrier density $n$ in fluorinated SLG and BLG. Solid orange triangles correspond to a SLG device (sample A) reported in Ref.~\onlinecite{Hong2012:PRL} with $n_{\rm F}=5\times 10^{11}\,\mathrm{cm}^{-2}$. This data was taken on the hole side. The BLG data, on the electron side, are shown in blue (W02), green (W38), and red (W03) squares. Open symbols indicate $T=1.7\,\mathrm{K}$, while solid symbols show the $T=0$ extrapolation $\tau_{\rm sat}^{-1}$. The gray dashed and dotted lines correspond to a power law dependence of $n^{-1}$, differing by a factor of two.}
\end{figure}

Figure~\ref{fig:expTauPhiOverTauM} plots the ratio of the dephasing rate over the momentum scattering rate $\tau_{\phi}^{-1}/\tau_{m}^{-1}$ as a function of carrier density $n$, combining current and prior data from fluorinated SLG and BLG samples \cite{Hong2012:PRL,Stabile2015}. Here, $\tau_{\rm m}^{-1} = {n e^2}/{\sigma m^{\star}}$ where $m^{\star}$ is the effective mass of bilayer graphene (see Fig.~\ref{fig:expMass} in Appendix~\ref{app:exp}). The collapse of all BLG data onto a single line, independent of $n_\mathrm{F}$, strongly indicates that both $\tau_{\phi}^{-1}$ and $\tau_{m}^{-1}$ originate from the fluorine adatoms. From the SLG trend line to the BLG trend line, the ratio $\tau_{\phi}^{-1}/\tau_{m}^{-1}$ increases by only a factor of 2, in spite of a $n_\mathrm{F}$ change of close to a factor of 10. These observations have inspired us to seek a unified theoretical framework that can capture the effects of fluorine adatoms in both charge and spin scattering sectors, and on both SLG and BLG in a self-consistent manner.

\section{\label{sec:model}Theory}
\begin{figure}
  \includegraphics[width=0.9\columnwidth]{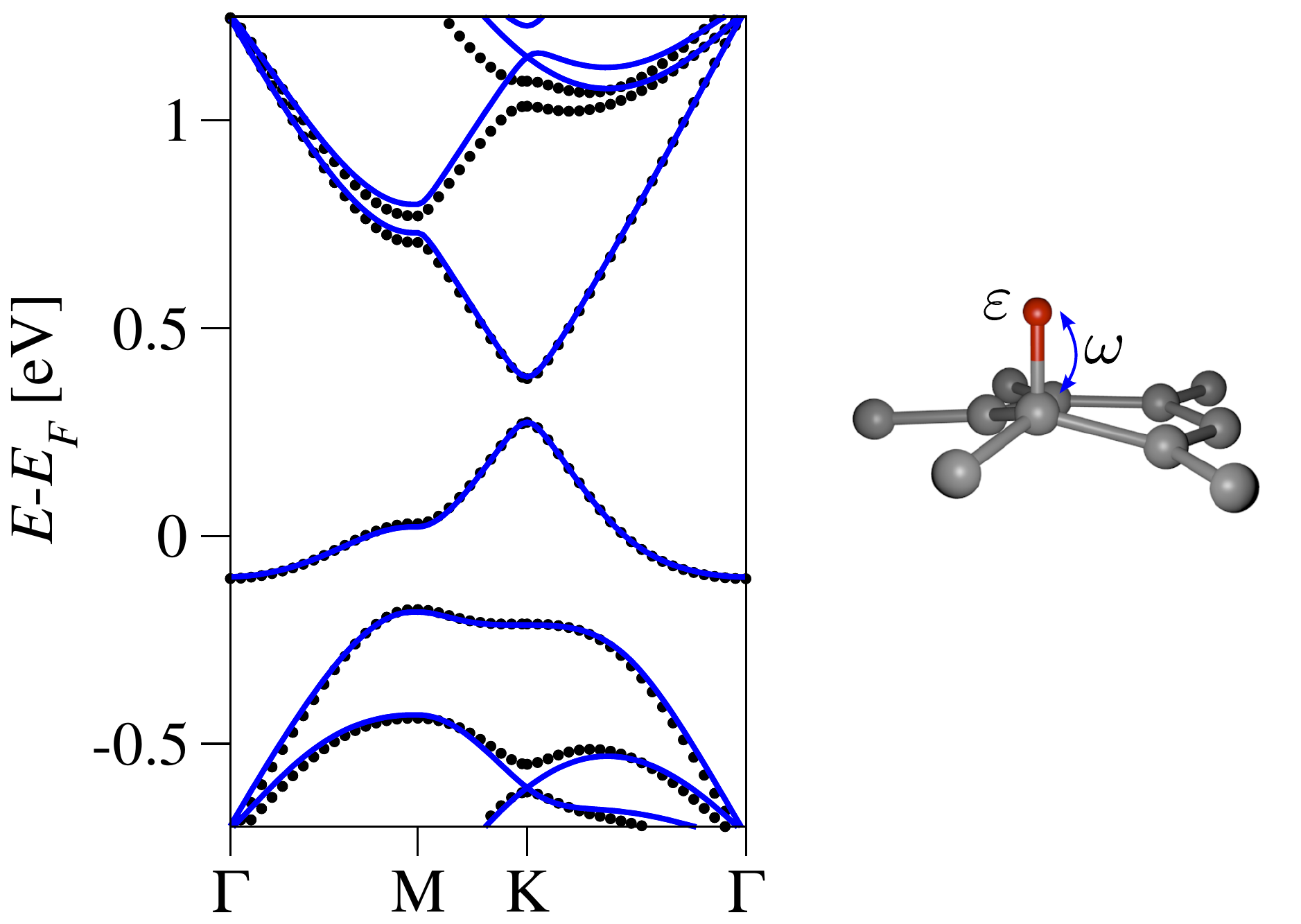}
  \caption{\label{fig:tbfit_fslg}DFT (black dotted) and TB (blue solid) calculated electronic band structure of a $10\times 10$ supercell of fluorinated SLG graphene. The TB parameters are $\omega=5.5\,$eV and $\varepsilon=-2.2\,$eV.}
\end{figure}
\begin{figure}
  \includegraphics[width=0.94\columnwidth]{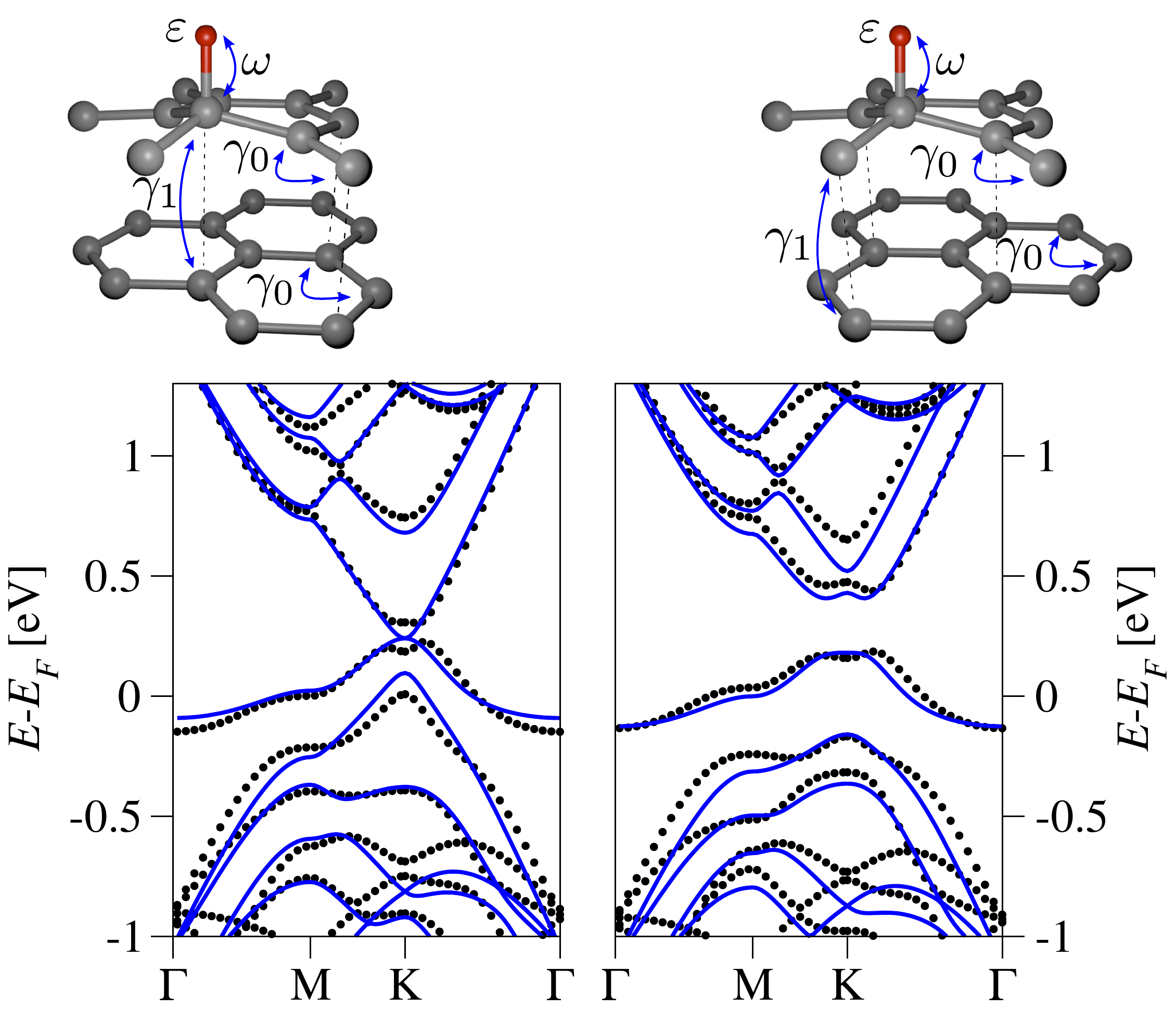}
  \caption{\label{fig:tbfit_fbil_u0}DFT (black dotted) and TB (blue solid) calculated electronic band structure of a $7\times 7$ supercell of BLG with one fluorine adatom in the dimer (left) and nondimer (right) adsorption position on the top layer. The TB parameters are $\omega_{\rm d}=7.0\,$eV and $\varepsilon_{\rm d}=-2.5\,$eV for the dimer configuration, and $\omega_{\rm nd}=8.0\,$eV, $\varepsilon_{\rm nd}=-3.0\,$eV for the nondimer configuration.}
\end{figure}

We model a single fluorine adatom in the TB approximation as an Anderson-like impurity carrying an effective local magnetic moment. The spin relaxation rates $\tau_{\rm s}^{-1}$ (and momentum relaxation rates $\tau_{\rm m}^{-1}$) are subsequently obtained through the fully analytical T-matrix approach~\cite{Kochan2014:PRL, Kochan2015:PRL}. The employed spin relaxation mechanism relies on resonant scattering off single impurities and the magnetic exchange interaction between the electron spin and the impurity-induced magnetic moment: An electron scattering resonantly off the impurity experiences a local spin-flip field. In case of long enough resonance lifetime, the electron spin is effectively randomized on exiting the scattering region. Details on the formalism can be found in Refs.~\onlinecite{Kochan2014:PRL,Kochan2015:PRL} and we also provide the analytic formulas in Appendix~\ref{app:tbmodel}.

Since DFT results are not conclusive~\cite{Santos2012:NJP, Kim2013:PRB} on whether fluorine carries a magnetic moment or not, we employ DFT and TB calculations to determine fluorine's orbital parameters and rely on the measured spin relaxation rates at $T=0$ in order to extract the exchange coupling $J$ due to a potential magnetic moment on fluorine. A reasonable $J$ value and good agreement between model and experiment would be a strong indicator that fluorine-induced magnetic moments are the dominant source for spin relaxation in this system.

The chemisorption of fluorine on top of SLG and AB stacked BLG is described by two TB parameters, the onsite energy $\varepsilon$ and the hybridization strength $\omega$ as schematized in Figs.~\ref{fig:tbfit_fslg} and \ref{fig:tbfit_fbil_u0}. The bare graphene structure is modeled in terms of the standard intralayer nearest-neighbor hopping $\gamma_0=2.6\,\mathrm{eV}$ and, in the BLG case, also by the interlayer coupling $\gamma_1=0.34\,\mathrm{eV}$. Two different adsorption positions, dimer (d) and nondimer (nd), are taken into account for BLG (see Appendix~\ref{app:tbmodel} for details). We assume that those positions are equally populated on the top layer during the fluorination process.

The orbital TB parameters for fluorine are obtained from fitting the spin unpolarized electronic band structure computed within DFT (see Appendix~\ref{app:dft} for details). We extract the TB parameters of $\varepsilon = -2.2\,\mathrm{eV}$ and $\omega = 5.5\,\mathrm{eV}$ for fluorinated SLG~\cite{Irmer2015:PRB}, and $\varepsilon_\mathrm{d} = -2.5\,\mathrm{eV}$, $\omega_\mathrm{d} = 7.0\,\mathrm{eV}$ and $\varepsilon_\mathrm{nd} = -3.0\,\mathrm{eV}$, $\omega_\mathrm{nd} = 8.0\,\mathrm{eV}$ for fluorinated BLG in the dimer and nondimer configurations, respectively. The fits are displayed in Figs.~\ref{fig:tbfit_fslg} and \ref{fig:tbfit_fbil_u0}. In contrast to the SLG, the fit is not unique in the BLG configuration. Charging effects between the two graphene layers are observed in the DFT data which seem to be responsible for the gap opening at the K point for the dimer position (see also Fig.~\ref{fig:tbfit_fbil_u16} in Appendix~\ref{app:tbmodel}). This gap is not recovered in our simple TB model. 

\begin{figure}
  \includegraphics[width=\columnwidth]{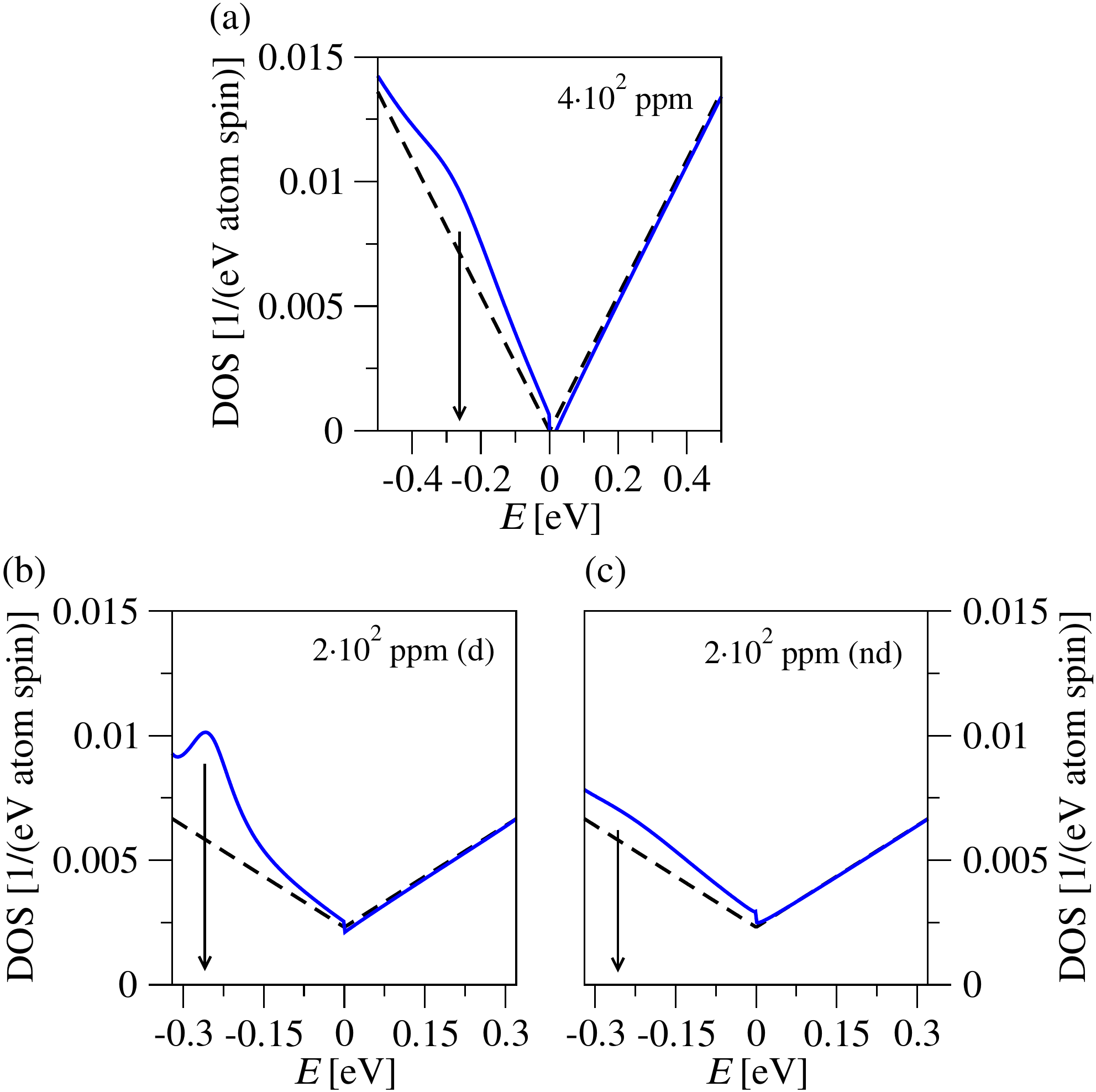}
  \caption{\label{fig:pdos_f_slbl_noJ} Perturbed DOS (blue solid) of fluorinated (a) SLG and (b)-(c) BLG. Compared to the DOS of pristine graphene (black dashed), broad ($200$-$300$\,meV) resonance levels appear due to fluorine at (a) $E_{\rm res} = -262\,$meV in SLG, (b) $E_{\rm res} = -253\,$meV in BLG with fluorine in the dimer position, and (c) $E_{\rm res} = -247\,$meV in BLG with fluorine in the nondimer position. For better visibility, the data are displayed for fluorine concentrations of $\eta = 400\,$ppm and $\eta = 200\,$ppm for SLG and BLG configurations, respectively. The arrows indicate the resonance positions.}
\end{figure}

Based on the orbital parameters extracted from the DFT calculations, we observe from the perturbed density of states (DOS), shown in Fig.~\ref{fig:pdos_f_slbl_noJ}, that fluorine induces broad resonances in graphene. The resonance levels lie at about $E_{\rm res}\approx -250\,$meV, which is significantly away from the charge neutrality point. For the formula of the perturbed DOS see Eq.~(\ref{eq:supp_pdos}) in Appendix~\ref{app:tbmodel}. As a consequence of the relatively large negative resonance energy, the calculated (spin and momentum) relaxation rates are expected to be significantly higher on the hole side~\cite{Irmer2018:PRB}. However, the measured $\sigma$ and $\tau_{\phi}^{-1}$ are roughly electron-hole symmetric.

% -------------------------------------------------------------------------------------------
\section{\label{sec:modExp}Results and comparison to experiment}
\begin{figure}
  \includegraphics[width=\columnwidth]{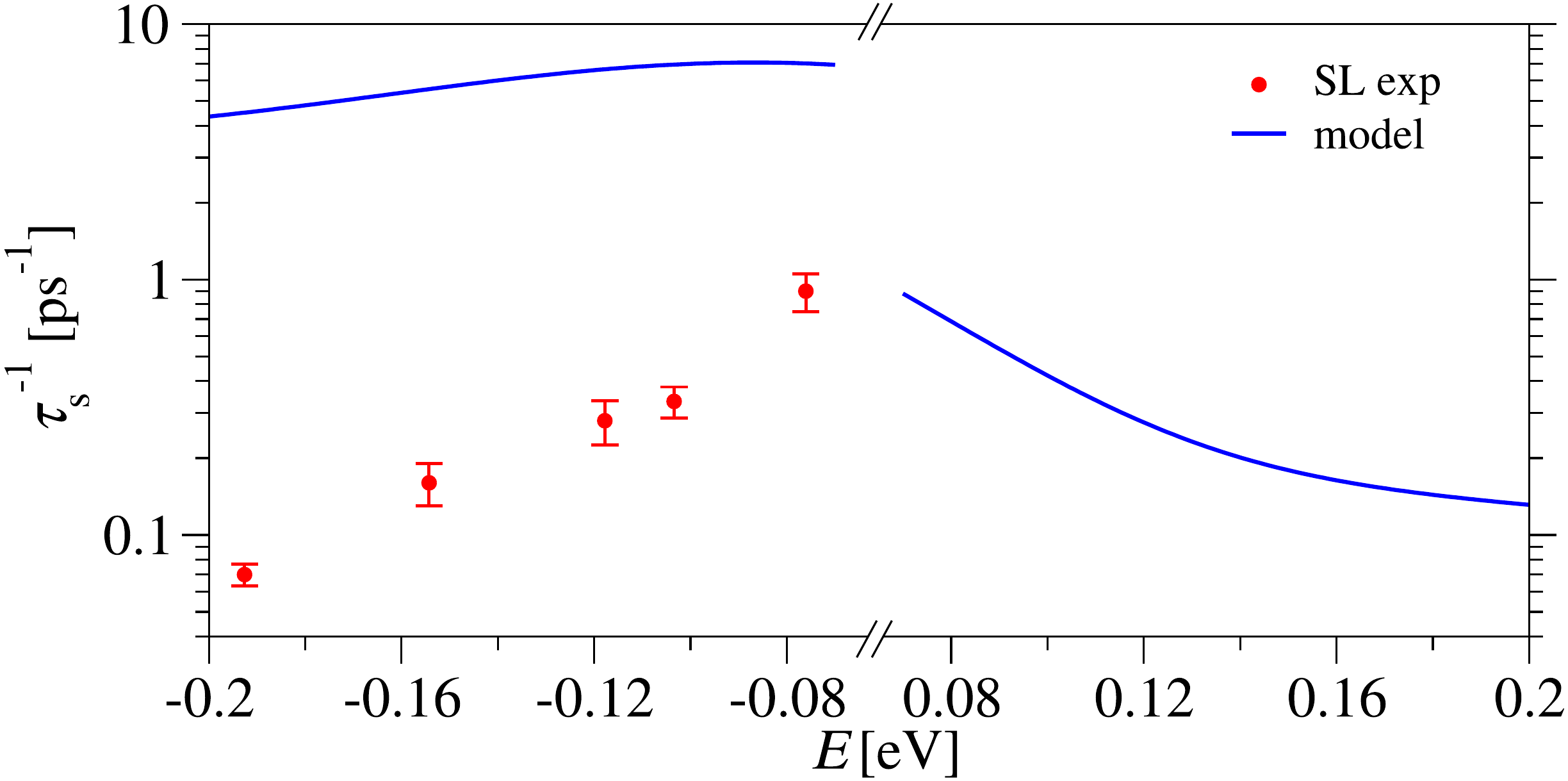}
  \caption{\label{fig:spinrel_sl_f} Spin relaxation rate in fluorinated SLG. The experimental data (red symbols) are from sample A of Ref.~\cite{Hong2012:PRL}. The calculations (blue solid) are obtained by taking fluorine's orbital parameters (see main text), setting $J = 0.56\,$eV, and employing a broadening of $\Sigma_{\rm eh}=64\,$meV due to electron-hole puddles. The concentration of fluorine adatoms is set to the experimental estimate of $\eta=131\,$ppm.}
\end{figure}

Figure~\ref{fig:spinrel_sl_f} plots the measured $\tau_{\rm sat}^{-1}$ and computed spin relaxation rate in fluorinated SLG for both electron and hole carriers. Fixing the fluorine concentration to the experimentally estimated value of $\eta = 131\,$ppm ($n_{\rm F}=5\times 10^{11}\,\mathrm{cm}^{-2}$) and taking into account a broadening $\Sigma_{\rm eh}$ of the rates due to electron-hole puddles, the only free parameter in our model is the exchange coupling $J$. On the hole side, no value of $J$ can be found to match the calculated $\tau_{\rm s}^{-1}$ to experimental data. To illustrate, we show the calculated $\tau_{\rm s}^{-1}$ using $J = 0.56\,$eV (and $\Sigma_{\rm eh}=64\,$meV consistent with experimental estimates \cite{Hong:PRB2009}). This model calculation produces $\tau_{\rm s}^{-1}$ of the experimental magnitude on the electron side but overshoots the measurement on the hole side by more than an order of magnitude. This comparison shows the discrepancy between model and measurement on the issue of electron-hole symmetry.

\begin{figure}
  \includegraphics[width=0.5\columnwidth]{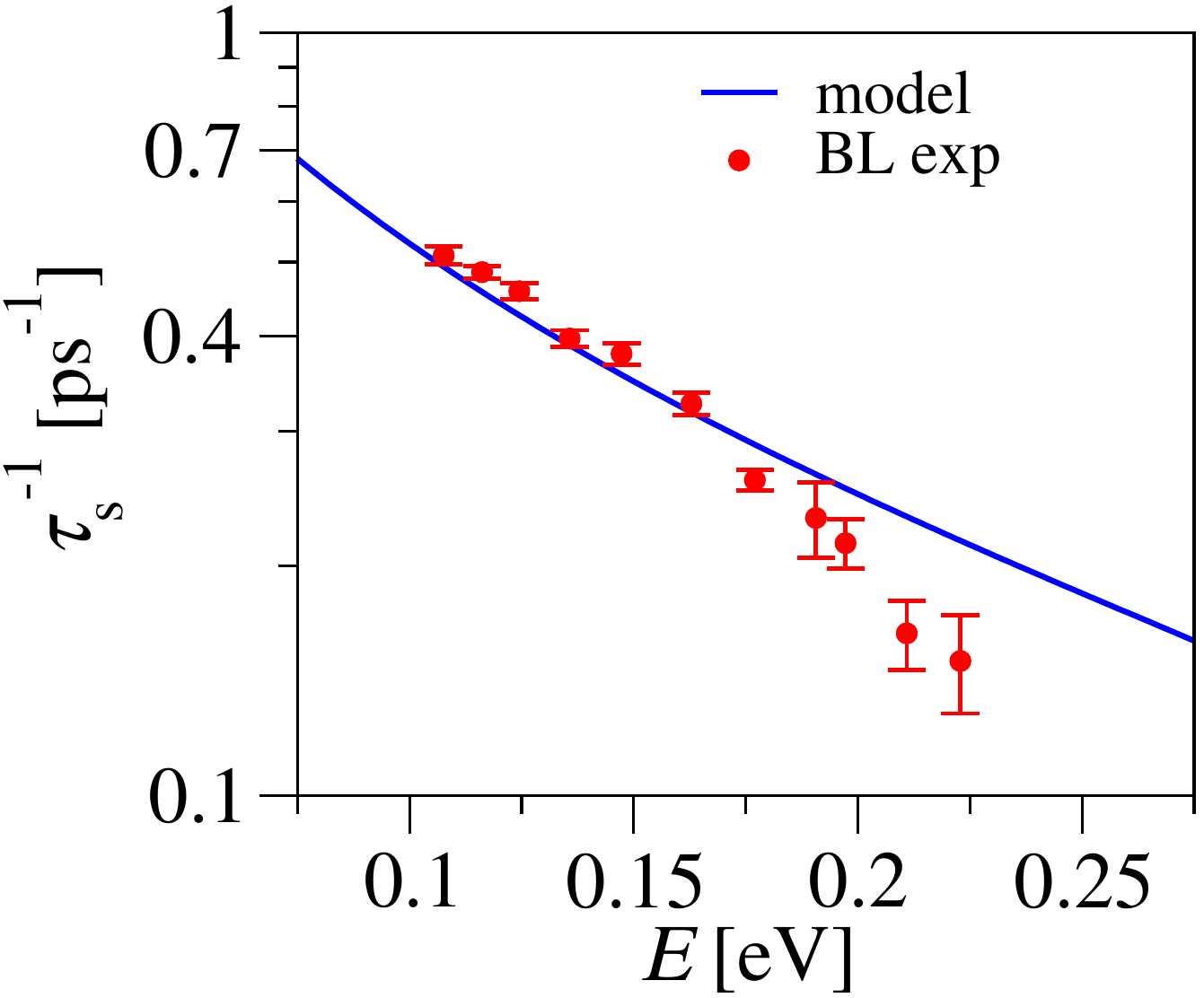}
  \caption{\label{fig:spinrel_bl_f}Spin relaxation rate in fluorinated BLG. The experimental data (red symbols) are from W03 at $T=0$. Calculations (blue line) are obtained by taking fluorine's orbital parameters (see main text), setting $J = 0.47\,$eV, and employing a broadening of $\Sigma_{\rm eh}=20\,$meV to the rates due to electron-hole puddles. The concentration of fluorine adatoms is set to the experimental estimate of $\eta=572\,$ppm.}
\end{figure}
Similar behavior is found when we compare $\tau_{\rm s}^{-1}$ of calculation and measurement in fluorinated BLG. An example using data for sample W03 ($n_{\rm F} = 4.4\times 10^{12}\,\mathrm{cm}^{-2}$) is shown in Fig.~\ref{fig:spinrel_bl_f}. In this case, calculations using $J = 0.47\,$eV and $\Sigma_{\rm eh} = 20\,$meV and the experimentally estimated fluorine concentration reproduce data very well, \emph{on the electron side}. However, the calculated $\tau_{\rm s}^{-1}$ is significantly higher on the hole side (not shown) whereas the measured $\tau_{\phi}^{-1}$ in Fig.~\ref{fig:expTauPhi} is manifestly electron-hole symmetric.

\begin{figure}
  \includegraphics[width=\columnwidth]{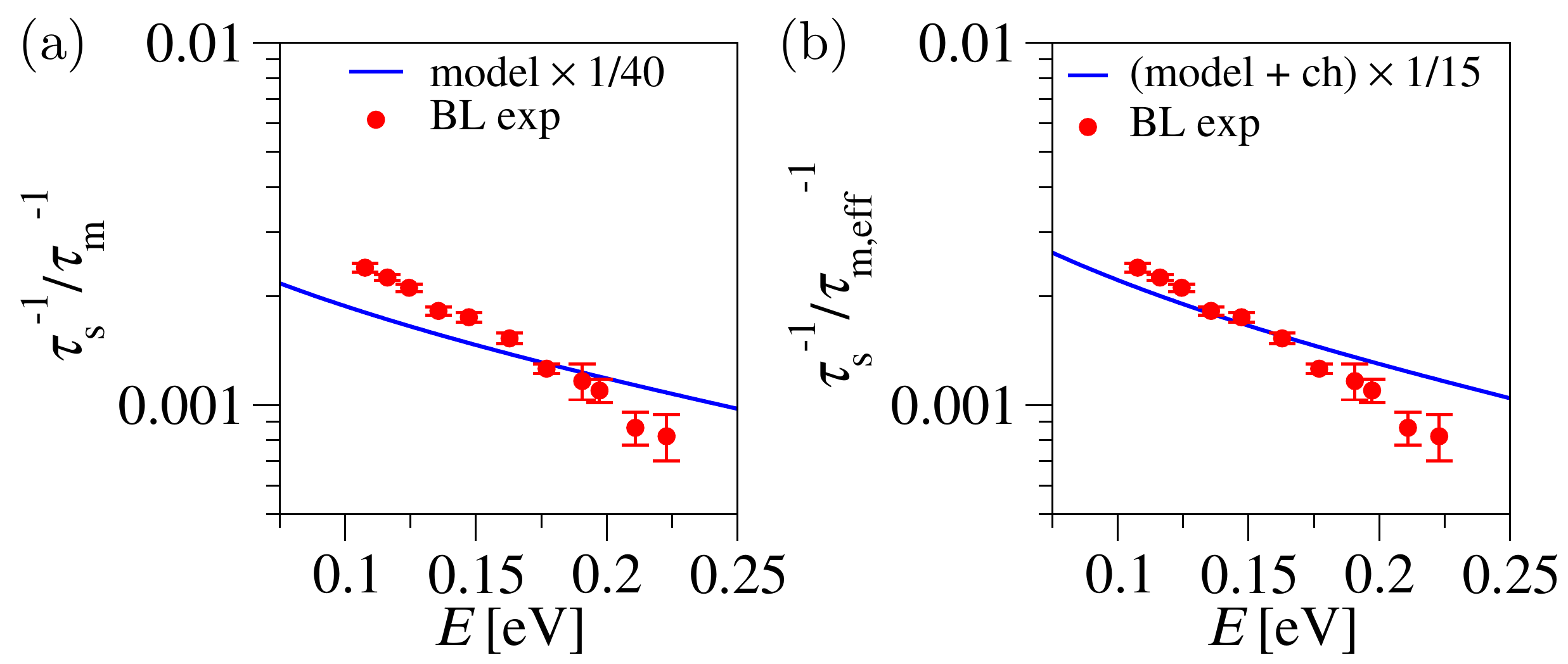}
  \caption{\label{fig:rateRatio_bl}Spin to momentum relaxation rate ratio of fluorinated BLG. (a) The model calculation (blue solid) overestimates the experimental values (red symbols, sample W03) by a factor of $40$. (b) Considering fluorine as a charged impurity reduces the discrepancy to a factor of $15$.}
\end{figure}
The evaluation of the ratio of spin to momentum relaxation rate, $\tau_{\rm s}^{-1}/\tau_{\rm m}^{-1}$, further exposes the difficulty of reconciling measurement and calculations based on the available DFT descriptions (Figs.~\ref{fig:tbfit_fslg} and \ref{fig:tbfit_fbil_u0}). Though Fig.~\ref{fig:spinrel_bl_f} shows a good agreement between the calculated and measured $\tau_{\rm s}^{-1}$ in fluorinated BLG on the electron side, the comparison of $\tau_{\rm s}^{-1}/\tau_{\rm m}^{-1}$ for the same range of data on the same sample shows that the model underestimates the experimental momentum relaxation rate by about a factor of 40, see Fig.~\ref{fig:rateRatio_bl}(a).
In other words, because the electron side is so far away from the resonance, the momentum scattering caused by fluorine adatoms is too weak to capture the measured conductivity.

Based on the observation of charging effects in the DFT data (see Appendix~\ref{app:tbmodel}), we have also treated each fluorine adatom as a scattering center carrying charge $-e$ and considered thus an upper bound for additional charged impurity scattering in this scenario. The details are given in Appendix~\ref{app:tbmodel}. The additional contribution, resulting in $\tau_{\rm m, eff}^{-1}$, reduced the data-calculation discrepancy to a factor of about $15$, as shown in Fig.~\ref{fig:rateRatio_bl}(b), which remains significant.

\begin{figure}
  \includegraphics[width=0.95\columnwidth]{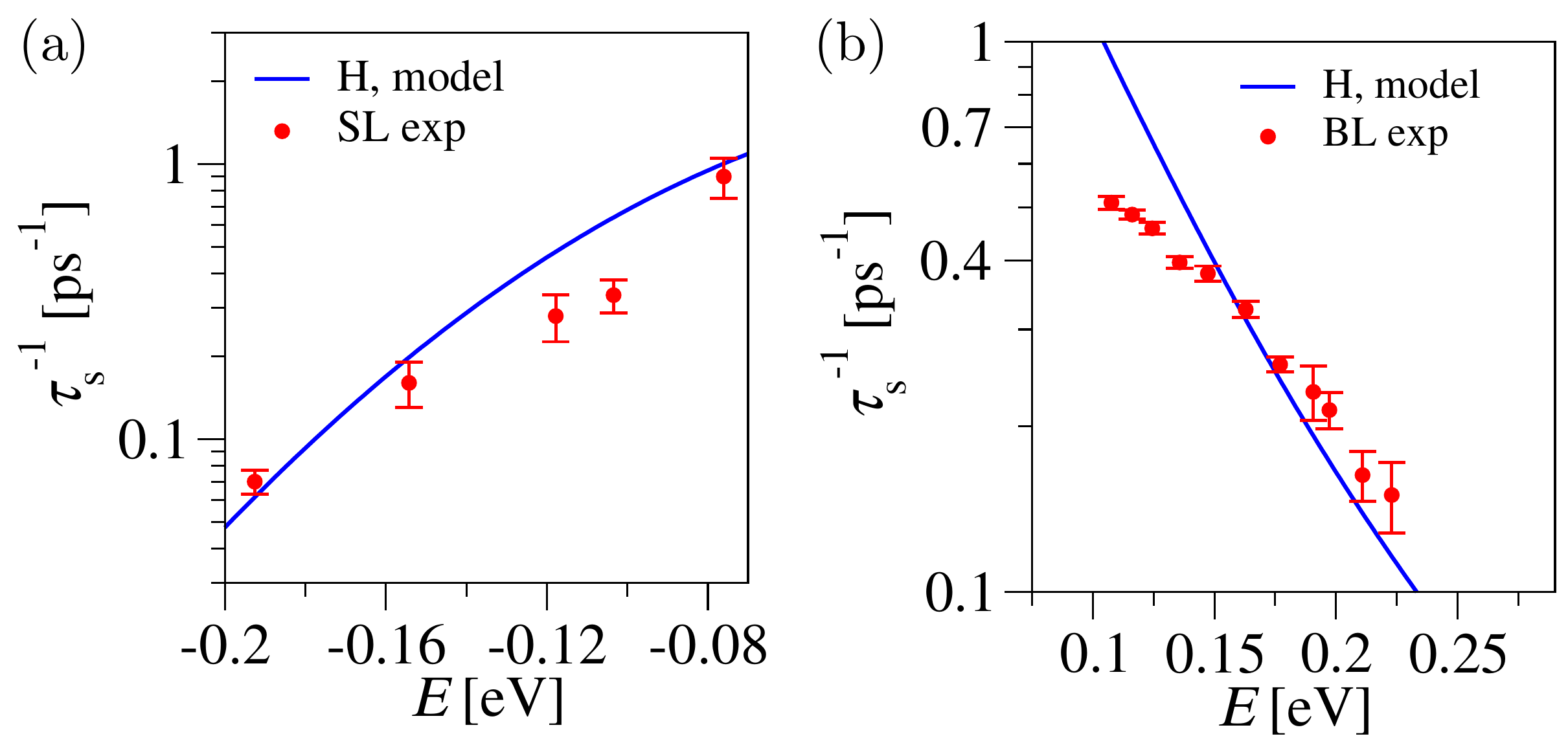}
   \caption{\label{fig:spinrel_slbl_h}Spin relaxation rate of fluorinated (a) SLG (sample A) and (b) BLG (sample W03). The experimental data (red symbols) can be described by hypothetical strong resonant impurities such as hydrogen (see parameters in the main text) as a source for spin relaxation. Setting the hydrogen concentration to the experimental estimates, (a) $\eta = 131\,$ppm and (b) $\eta = 572\,$ppm, exchange couplings of $J = 9\,$meV and $J = -40\,$meV are extracted for the SLG and BLG case, respectively. The rates were broadened by (a) $\Sigma_{\rm eh}=77\,$meV and (b) $\Sigma_{\rm eh}=20\,$meV.}
 \end{figure}

The above comparisons point to a key difference between experiment and DFT calculations, i.e., fluorine appears to behave like a strong resonant impurity in experiment while its DFT description is clearly not. Below we offer a few thoughts on what can cause the disagreement, with the hope to stimulate further studies. 

We considered the possibility of small fluorine clusters ($<2\,$nm) as clusters may give rise to symmetric conductivity~\cite{Katsnelson2009,McCreary2010}. Clustering simultaneously quenches the Raman signal~\cite{Lucchese2010} of an isolated adatom, reduces its resonant impurity scattering strength~\cite{Katsnelson2009,McCreary2010}, and most likely quenches potential magnetic moments~\cite{Nair2012:NP,Wilhelm2015} which would affect both the spin relaxation rate and the scattering rate ratio. We can not rule out their presence in our samples, but clustering can not solve the $\tau_{\rm s}^{-1}/\tau_{\rm m}^{-1}$ puzzle in our opinion. A quantitative evaluation could shed further light on the role of clustering.

Another important factor is lattice deformation. Experimentally we have noticed that the presence of local curvatures, e.g. created by exfoliating to a rough substrate such as SiO$_{2}$, is essential to the fluorination process. This suggests that the local bonding and ionic environment of a fluorine adatom in real devices is likely quite different from that of a DFT simulation. A realistic description of the adatoms may be crucial to capture their electronic properties accurately. This can potentially reconcile the difference between the DFT electronic structure of fluorine adatoms appearing as weak resonances off the 
charge neutrality point and the experimental indication of fluorine being a midgap scatterer. More elaborate DFT studies would be needed to confirm this hypothesis.

Though not supported by our current DFT calculations, the quantitative similarities between the conductivity measurements of fluorinated~\cite{Hong2012:PRL} and hydrogenated~\cite{Ni2010} graphene motivated us to model fluorine as a hypothetical strong resonant impurity, which induces resonance levels very close to the charge neutrality point and thus can produce the experimentally observed electron-hole symmetry~\cite{Irmer2018:PRB}. We used here the orbital parameters for hydrogen, $\varepsilon=0.16\,$eV, $\omega=7.5\,$eV for SLG~\cite{Gmitra:PRL2013}, and $\varepsilon_{\rm d}=0.25\,$eV, $\omega_{\rm d}=6.5\,$eV, $\varepsilon_{\rm nd}=0.35\,$eV, $\omega_{\rm nd}=5.5\,$eV for BLG~\cite{Kochan2015:PRL}. Fits to the experimental $\tau_{\rm s}^{-1}$ (with fixed $n_{\rm F}$) led to small exchange strengths of $J=9\,$meV ($\Sigma_{\rm eh}=77\,$meV) in the SLG case, and $J=-40\,$meV ($\Sigma_{\rm eh}=20\,$meV) in the BLG case.

\begin{figure}
  \includegraphics[width=\columnwidth]{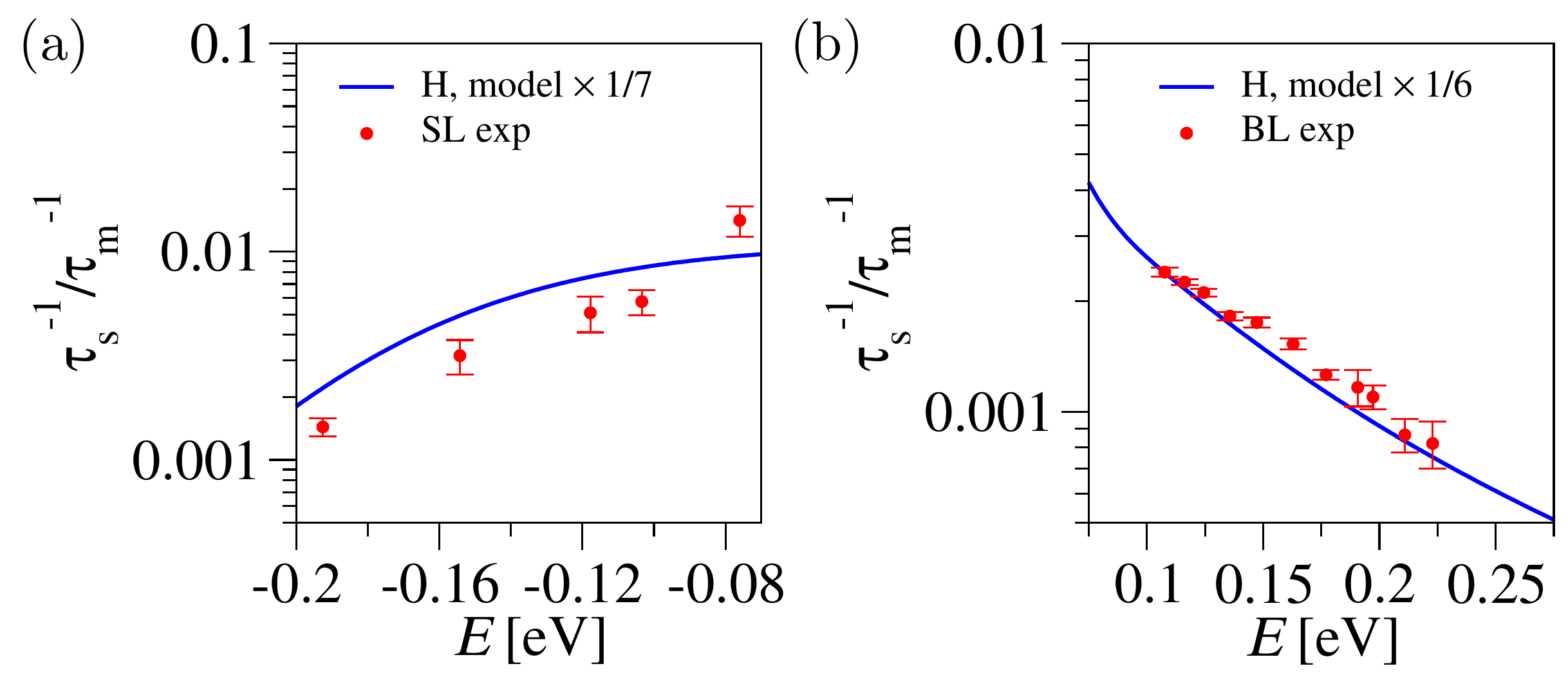}
  \caption{\label{fig:rateRatio_slbl_h}Spin to momentum relaxation rate ratio of fluorinated (a) SLG (sample A) and (b) BLG (sample W03) modeled by hypothetical strong resonant impurities such as hydrogen (blue solid). Calculations overestimate the experimental values (red symbols) by a factor of (a) $7$ and (b) $6$.}
\end{figure}

Figure~\ref{fig:spinrel_slbl_h} shows that the experimental spin relaxation rates in both SLG and BLG (holes and electrons) can be reproduced well with these parameters. which is a significant improvement compared to Fig.~\ref{fig:spinrel_sl_f}. Examining the ratio of the spin and momentum scattering rates, we still observe an underestimation of the momentum relaxation rate in both SLG and BLG samples by about a factor of $7$ and $6$, respectively, as shown in Fig.~\ref{fig:rateRatio_slbl_h}. The underestimation is consistent with a previous study, where we showed that even strong resonant adatoms such as H do not produce momentum relaxation rates as high as a vacancy or strong midgap scatterer does\cite{Irmer2018:PRB}. Experimental data of charge scattering in fluorinated and hydrogenated graphene~\cite{Hong2012:PRL, Ni2010}, on the other hand, seem to fit the scattering model of a strong midgap scatterer rather well~\cite{Stauber2007,Ferreira:PRB2011,Monteverde2010,Robinson2008}. This is another puzzling aspect of functionalized graphene that needs to be understood before quantitative assessments of scattering processes can be accurately made.

The relatively small values of the exchange coupling $J$ obtained in the fitting of Fig.~\ref{fig:spinrel_slbl_h} suggest that spin-flip scatterings caused by fluorine adatoms are weak. As mentioned in the introduction, the induction of a magnetic moment in fluorinated graphene is quite subtle and depends on a set of other parameters such as doping and fluorine concentration. Furthermore, a recent study~\cite{Jiang:NatComm2018} showed that vacancy-induced magnetic moments in graphene can be screened by itinerant electrons, where the Kondo temperature depends on gating and the local curvature of the graphene sheet. Should similar physics occur for fluorine, a fraction of the fluorine-induced moments may be screened and manifests as a reduced exchange coupling $J$ in our fittings.

Other possibilities include correlated-impurity effects currently not evaluated in our spin relaxation model and possibly new phase breaking mechanisms that are nonmagnetic in origin, that could complicate the WL data analysis. In this regard, it is worth mentioning that calculations have shown that SOC terms which preserve the mirror symmetry of the graphene plane, i.e. $S_z$, can lead to spin-dependent scattering that mimics the effect of spin-flip scattering in the WL measurements~\cite{McCann:PRL2012}. Fluorine induces a local SOC of about $10\,$meV~\cite{Irmer2015:PRB}. However, according to earlier studies in Ref.~\onlinecite{Bundesmann:PRB2015} on fluorinated SLG, an approximately one thousand times higher concentration would be necessary to reach the measured dephasing rate. This suggests that the local SOC induced by fluorine is not the dominant source of spin relaxation observed in experiment.

\section{\label{sec:conclude}Concluding remarks}
In conclusion, we performed a comprehensive experiment-theory study to investigate the effect of a dilute concentration of fluorine adatoms on both the spin and charge relaxation of carriers in single-layer and bilayer graphene. Experimental evidence points to fluorine being the dominant source of both spin and momentum scattering. In the charge channel, fluorine behaves as a strong midgap scatterer that is situated at the charge neutrality point whereas in the spin channel, 
the experiment suggests it is a weak spin-flip scatterer. 

Theoretically, we calculated supercell electronic structures of graphene with fluorine adatoms and obtained tight-binding models for investigating scattering by the T-matrix formalism. We present spin and momentum relaxation rates in the limit of independent dilute magnetic scatterers. The 
modeling predicts fluorine-induced resonances off the neutrality points, leading to a marked difference between electron and hole transport channels. This is at odds with the experiment. Also, 
the model predicts a rather strong spin flip scattering and weaker momentum relaxation rates than 
the measured data. The agreement with the experiment cannot be reconciled by considering 
charged adatoms (due to charge transfer between fluorine and graphene), nor by reducing the exchange
coupling. However, the agreement improves significantly if we use a hydrogen model, which is 
very close to the midgap scatterer model, yielding only weakly electron-hole asymmetric results.
This model still underestimates the momentum relaxation rate by a few fold.

The comparison between experiment and theory highlights practical complications and challenges that need to be overcome before the electronic properties of the fluorine adatom, a widely used functionalization element on two-dimensional materials, can be accurately captured in DFT calculations. There is still a profound lack of understanding (and agreement) on the presence
or absence of a magnetic moment on dilute fluorine adatoms on graphene. From our study we also 
see that the basic electronic structure obtained from DFT can miss significant practical sample
features, such as the structural deformations discussed above. The original data
and rather deep theoretical analysis into the current state of knowledge about the system should
provide further impetus to investigate the fascinating physics of resonant scattering and 
spin relaxation in graphene functionalized not only with fluorine, but also other types of adatoms.

\appendix

\section{\label{app:exp}Experimental dephasing and saturation rate}
\begin{figure}
  \includegraphics{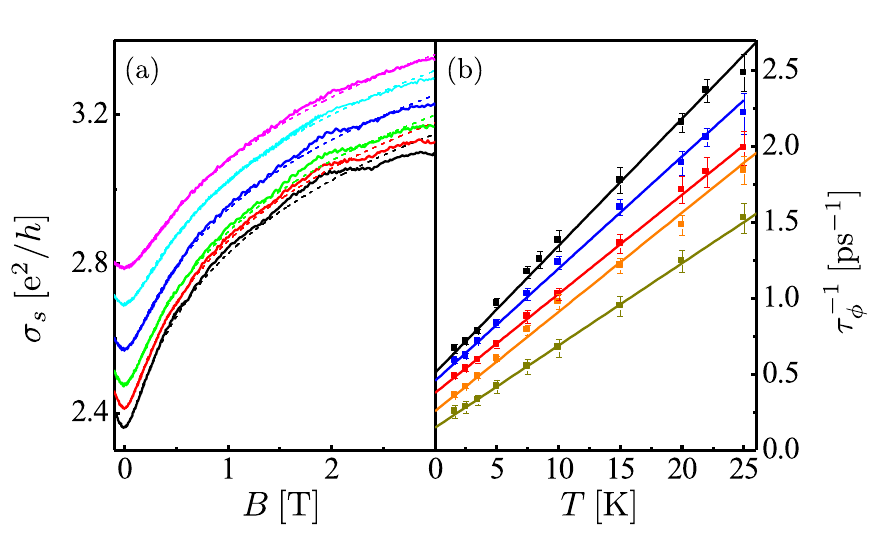}
    \begin{minipage}{0.7\columnwidth}
 \begin{ruledtabular}
    \begin{tabular}{l|c|c|c|c|c}
      $\mathrm{n}\:[10^{12}\,\mathrm{cm}^{-2}]$ & 5 & 6 & 7.5 & 9.5 & 13 \\\hline
      $\alpha$ & 1.81 & 1.63 & 1.51 & 1.63 & 1.56
    \end{tabular}
  \end{ruledtabular}
  \end{minipage}
  \caption{\label{fig:expSheetCondB}(a) Sheet conductance $\sigma_s$ vs $B$ for BLG sample W03 measured at electron density $n=6\times 10^{12}\,\mathrm{cm}^{-2}$. From bottom to top: $T=1.6$, 2.5, 3.5, 5.0, 7.5, 10\,K. Dashed lines are fits to the WL expression in BLG of Ref.~\onlinecite{Gorbachev:PRL2007}. (b) Dephasing rate $\tau_{\phi}^{-1}$ in sample W03 as a function of temperature at varying electron densities $n$. From bottom to top: $n=13$, 9.5, 7.5, 6.0, $5.0\times 10^{12}\,\mathrm{cm}^{-2}$. The table presents the extracted coefficient $\alpha$ in Eq.~(\ref{eq:alpha}).}
\end{figure}

\begin{figure}
  \includegraphics[width=0.7\columnwidth]{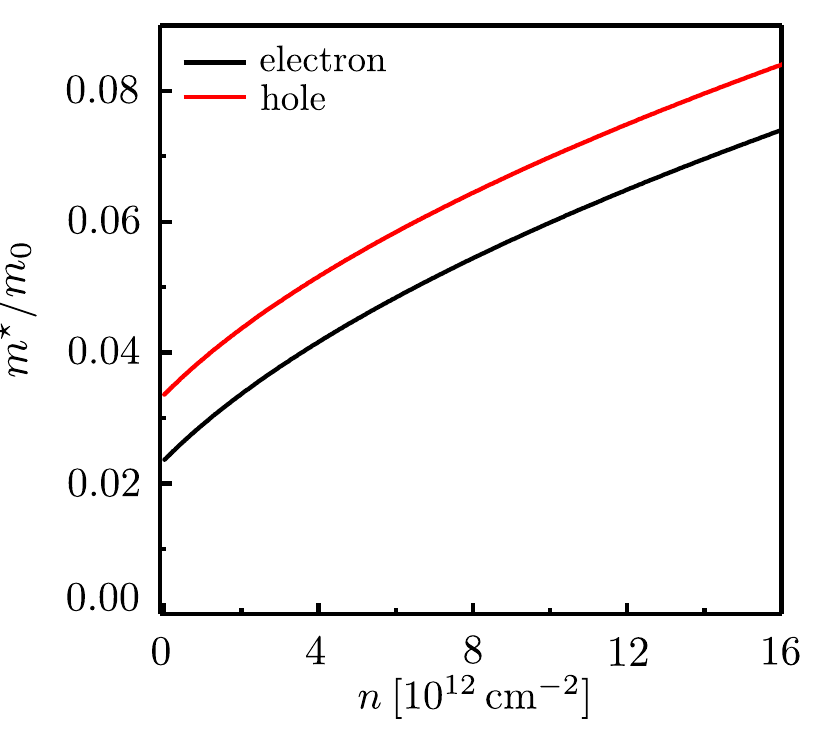}
  \caption{\label{fig:expMass}Experimental effective mass of electrons and holes in BLG using tight-binding parameters $\gamma_0=3.43\,\mathrm{eV}$, $\gamma_1=0.40\,\mathrm{eV}$, $\gamma_4=0.216\,\mathrm{eV}$, $\gamma_3=0$, $\Delta=0.018\,\mathrm{eV}$ as obtained in Refs.~\onlinecite{Zou:PRB2011,Li:PRB2016}.}
\end{figure}

Figure~\ref{fig:expSheetCondB}(a) plots the magnetoconductance $\sigma_{s}(B)$ of W03 at $n_e = 6\times 10^{12}/\mathrm{cm}^2$ and selected temperatures. Fits to Eq.~(1) of Ref.~\onlinecite{Gorbachev:PRL2007} are shown as dashed lines and provide an excellent description of data. The phase decoherence length $l_{\phi}$ obtained from the fits ranges from 30 to 114\,nm, from which we obtain the dephasing rate $\tau_{\phi}^{-1}$ through $\tau_{\phi}^{-1} = D_i l_{\phi}^{-2}$ where $D_i$ is the diffusion constant given by
\begin{equation}%
  D_i=\frac{\sigma_d h^{2}}{8\pi m^{\star}e^2}.
\end{equation}%
Here, $\sigma_d$ is the Drude sheet conductance measured around $T=200\,$K and $m^\star$ the $n$-dependent effective mass of BLG calculated for the current density range using experimentally determined TB parameters \cite{Zou:PRB2011,Li:PRB2016}. The values of $m^{\star}$ are given in Fig.~\ref{fig:expMass}. The fits also use $l_i = l^{\star} =10\,\mathrm{nm}$ although varying $l_i$ and $l^\star$ by a factor of two up or down has negligible effect on $l_\phi$ which is given by the low magnetic field regime ($B \lesssim 0.5$\,T). The values of $l_i$  and $l^\star$  are roughly the inter-fluorine spacing, similar to what we found on fluorinated SLG \cite{Hong2012:PRL}. Similar measurements and analyses are performed up to $T = 35\,\mathrm{K}$ and at electron densities $n$ ranging from $5\times 10^{12}\,\mathrm{cm}^{-2}$ to $1.3 \times 10^{13}\,\mathrm{cm}^{-2}$. Figure~\ref{fig:expSheetCondB}(b) plots the resulting $\tau_{\phi}^{-1}(T)$ at different carrier densities. It is clear from the plot that $\tau_{\phi}^{-1}(T)$ follows a linear trend given by $\tau_{\phi}^{-1}=aT+\tau_{\rm sat}^{-1}$, with the slope $a$ ranging from $0.05$-$0.08\,\mathrm{ps}^{-1}/\mathrm{K}$.  We attribute the $aT$ term to electron-electron collision induced dephasing. It can be further written as
\begin{equation}%
a=\alpha k_B\frac{\ln g}{\hbar g},\label{eq:alpha}
\end{equation}%
where $g =  \sigma_d h/e^2$ is the dimensionless Drude sheet conductance. The resulting $\alpha$ ranges between 1.5 and 1.8 (see the table in Fig.~\ref{fig:expSheetCondB}), in excellent agreement with previous WL studies in pristine SLG \cite{Tikhonenko:PRL2008}, BLG \cite{Gorbachev:PRL2007} and our fluorinated SLG samples \cite{Hong2012:PRL}. The $T=0$ dephasing rate $\tau_{\rm sat}^{-1}$ of sample W03 is used to compare to calculations.

\section{\label{app:dft}DFT calculation}
The electronic structure of fluorinated BLG has been calculated within the DFT\cite{Hohenberg1964:PR} using the plane wave pseudopotential code Quantum ESPRESSO\cite{Giannozzi2009:JPCM}. A $10\times 10$ supercell for fluorinated SLG and $7\times 7$ supercell of Bernal stacked BLG in a slab geometry with a vacuum spacing of 15~${\rm\AA}$ were considered. The reduced Brillouin zone was sampled with $10\times 10$ $k$-points. The atomic positions in the supercell calculations have been relaxed using the quasi-newton algorithm based on the trust radius procedure. For the atomic species we have used projector augmented-wave pseudopotentials\cite{Kresse1999:PRB} with the PBE exchange-correlation functional\cite{Perdew1996:PRL} with kinetic energy cut-offs of 50~Ry for the wave function and 350~Ry for the density. The supercell sizes were chosen such that interference effects between the periodic images of fluorine in the supercell approach can be neglected. Therefore, we take these calculations as a reliable basis for our TB model which we will employ for the experimental measurements of dilute fluorinated graphene. 

\section{\label{app:tbmodel}Model}

\paragraph{Tight-binding model:}We describe fluorine in SLG and BLG\cite{Kochan2014:PRL,Kochan2015:PRL} as an Anderson-like impurity that possesses a non-itinerant magnetic moment. In the BLG case, we distinguish whether fluorine adsorbs on the dimer or nondimer carbon site, $\rm {C_d}$ and $\rm{C_{nd}}$, respectively. Within the TB approximation the full model Hamiltonian reads
\begin{equation}
H = H_0(\gamma_0,\gamma_1) + H'(\varepsilon,\omega) + H_\mathrm{ex}(J)\,.\label{eq:total_hamiltonian}
\end{equation}
The Hamiltonian $H_0$ describes unperturbed SLG or BLG with the nearest-neighbor intralayer hopping $\gamma_0$ and the interlayer hopping $\gamma_1$ ($\gamma_1 = 0$ in the SLG case), $H'$ describes the fluorine chemisorption with the onsite energy $\varepsilon$ and hybridization strength $\omega$, and $H_\mathrm{ex}$ represents the exchange interaction term with coupling $J$.
In more detail, for SLG we have
\begin{align}
H_{0}^\mathrm{SL}&=-\gamma_0\sum_{\langle m,n\rangle \sigma} |a_{m\sigma}\rangle\,\langle b_{n\sigma}| + {\rm h.c.} \label{eq:SL-hamiltonian}\,,
\end{align}
and for AB-stacked BLG
\begin{align}
H_{0}^\mathrm{BL}&=-\gamma_0\sum_{\langle m,n\rangle \sigma \atop \lambda\in\{\mathrm{t},\mathrm{b}\}}|a_{m\sigma}^{\lambda}\rangle\,\langle b_{n\sigma}^{\lambda}|
+\gamma_1\sum\limits_{m\sigma} |a_{m\sigma}^{\mathrm{t}}\rangle\,\langle b_{m\sigma}^{\mathrm{b}}| + {\rm h.c.} \label{eq:BL-hamiltonian}\,,
\end{align}
where $\gamma_0 = 2.6\,\mathrm{eV}$ and $\gamma_1= 0.34\,\mathrm{eV}$. Our AB-stacking assumes that $\gamma_1$ connects the sublattice A of the top ($\lambda = \rm t$) and the sublattice B of the bottom layer ($\lambda = \rm b$), respectively.
A carbon $2p_z$ orbital with spin $\sigma$, which resides on the lattice site $m$, is represented by the one-particle state $|c_{m\sigma}\rangle$, where $c=\{a,b\}$ depends on the sublattice degree of freedom of the site $m$. Similarly, $|f_{\sigma}\rangle$ stands for the fluorine $2p_z$ orbital with spin $\sigma$.
The fluorine adsorption is characterized by the two orbital TB parameters---the onsite energy $\varepsilon$ and the hybridization strength $\omega$:
\begin{equation}\label{eq:perturbation-hamiltonian}
H'=\varepsilon\sum\limits_{\sigma}|f_{\sigma}\rangle\,\langle f_{\sigma}|+\omega\sum\limits_{\sigma}
\left(|f_{\sigma}\rangle\,\langle c^*_{\sigma}|+\mathrm{h.c.}\right)\,,
\end{equation}
where $|c^*_{\sigma}\rangle$ denotes a carbon orbital that bonds with fluorine. To distinguish SLG and BLG cases, we use $\varepsilon$ and $\omega$ without any subscripts for the former case and we add the subscripts d and nd for the dimer and nondimer BLG positions, respectively.
We extract the orbital parameters $\varepsilon$ and $\omega$ by fitting the TB model Hamiltonian $H_0+H'$ to DFT data for spin unpolarized electronic band structures of fluorinated SLG and BLG, respectively. The resulting parameters are given in Sec.~\ref{sec:model}.

Fluorine's local magnetic moment is captured by the exchange term in Eq.~(\ref{eq:total_hamiltonian}),
\begin{equation}
H_\mathrm{ex} = -J\,\hat{\mathbf{s}}\cdot\hat{\mathbf{S}}\,.
\end{equation}
The energy-independent exchange strength $J$ couples the itinerant electron spin with the localized impurity spin (spin $\nicefrac{1}{2}$) being represented by the array of Pauli matrices $\hat{\mathbf{s}}$ and $\hat{\mathbf{S}}$, respectively.

\paragraph{\label{app:dft_charging}Charging effect:}

\begin{figure}
  \includegraphics[width=\columnwidth]{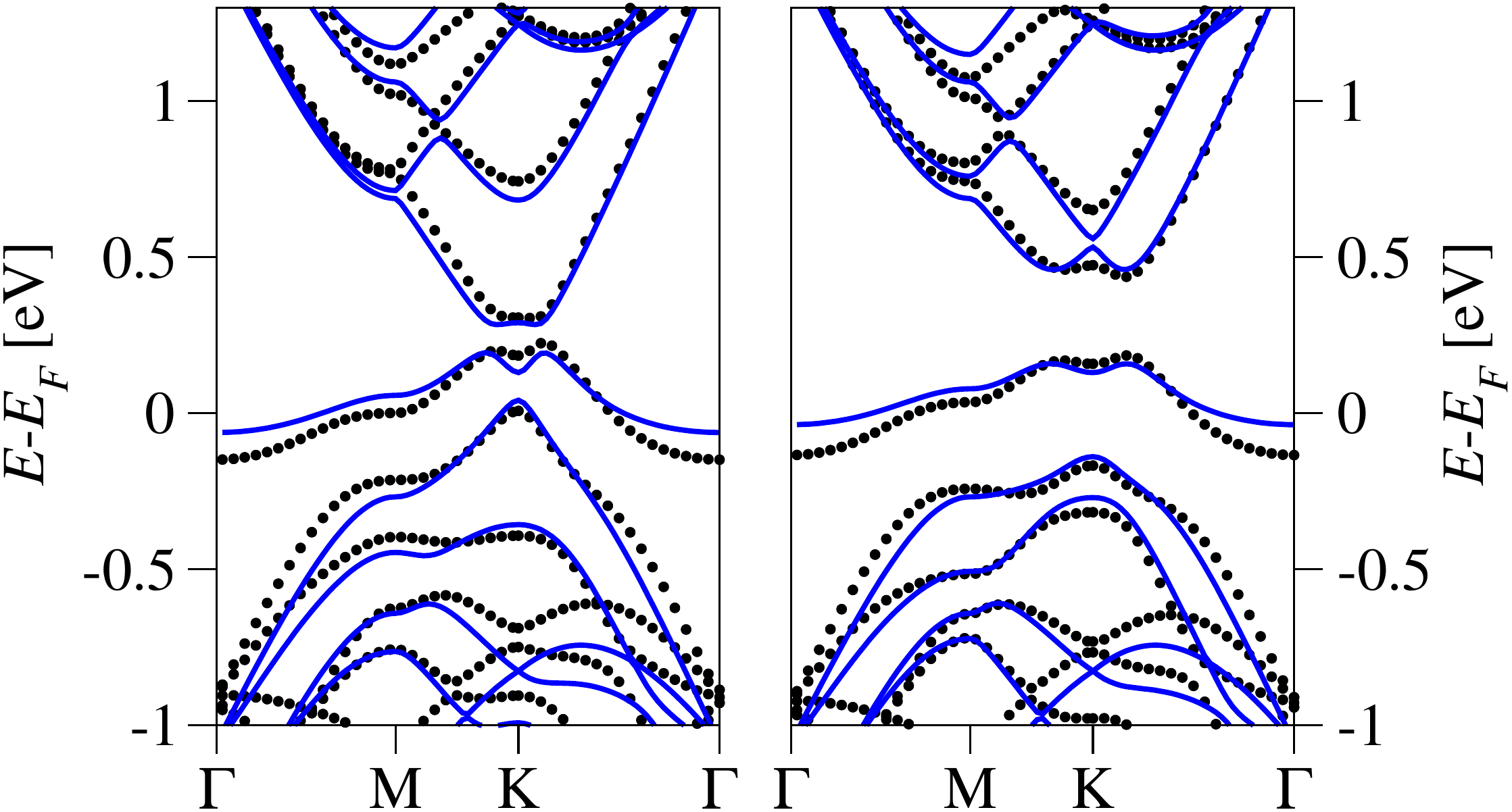}
  \caption{\label{fig:tbfit_fbil_u16}DFT (black dotted) and TB (blue solid) calculated electronic band structure of a $7\times 7$ supercell of BLG with one fluorine adatom in the dimer (left) and nondimer (right) adsorption position on the top layer. Additionally to the TB parameters $\omega_{\rm d}=7.0\,$eV and $\varepsilon_{\rm d}=-2.5\,$eV for the dimer configuration, and $\omega_{\rm nd}=8.0\,$eV, $\varepsilon_{\rm nd}=-3.0\,$eV for the nondimer configuration, a finite potential offset of $U=0.16\,$eV is assigned to the top layer of BLG in the TB calculation to account for charging effects.}
\end{figure}

The simple TB model above does not reproduce the gap opening between the two bands just above the Fermi level at the K point in the dimer configuration, see Fig.~\ref{fig:tbfit_fbil_u0}. Though, by adding a potential offset to $H_0$ which raises the onsite energies of all $2p_z$ orbitals on the upper layer, i.e.~$H_0+U\sum_{m,\sigma} |c^t_{m\sigma}\rangle\,\langle c^t_{m\sigma}|$, we can qualitatively improve the matching of the two considered electronic band structures. Fitting $U$ we found $U = 0.16\,\mathrm{eV}$; see Fig.~\ref{fig:tbfit_fbil_u16}. We attribute this potential offset to charging effects: The high electronegativity of the fluorine leads to charge redistribution among the BLG sheets. This potential offset is ignored in the calculation of the spin relaxation rates. The DFT calculations predict also for fluorinated SLG a charge transfer from graphene to fluorine~\cite{Irmer2015:PRB}, though no modification of the TB model is needed to reproduce the band structure.

\paragraph{Relaxation rates for resonant impurity scattering:}

The relaxation rates are computed from the underlying TB model, Eq.~(\ref{eq:total_hamiltonian}), employing the fully non-perturbative T-matrix approach~\cite{Kochan2014:PRL,Kochan2015:PRL}. The impurity's spin degrees of freedom double the one-particle state basis $|c_{m\sigma}\rangle,|f_\sigma\rangle\rightarrow|c_{m\sigma}\rangle\otimes|\Sigma\rangle,|f_\sigma\rangle\otimes|\Sigma\rangle$, where $\Sigma=\{\uparrow,\downarrow\}$ stands for the component of the impurity spin along the quantization axis. Introducing singlet ($\ell = 0$) and triplet 
($\ell = 1$) spin states and downfolding the Hamiltonian by decimating the $|f\rangle$ degrees of freedom, one obtains an analytic expression for the 
T-matrix~\cite{Kochan2014:PRL,Kochan2015:PRL} $\mathrm{T}(E)=\sum_{\ell,m_\ell} \mathrm{T}_\ell(E)\ |c_{\ell,m_\ell}\rangle\,\langle c_{\ell,m_\ell}|$, where
\begin{align}
\mathrm{T}_\ell(E)=\tfrac{{V}_\ell(E)}{1-{V}_\ell(E)\,G_{\mathrm{C}}(E)}\,,\ \ \
{V}_\ell(E)=\tfrac{\omega^2}{E-\varepsilon+(4\ell-3)J}\,.
\end{align}
The T-matrix contains the Green's function, $G_{\mathrm{C}}(E)=\langle c^*_{\uparrow}|(E+i\delta-H_0)^{-1}|c^*_{\uparrow}\rangle=\langle c^*_{\downarrow}|(E+i\delta-H_0)^{-1}|c^*_{\downarrow}\rangle$, of the unperturbed SLG or BLG which is projected to the carbon atomic site C hosting the fluorine adatom. In detail, $G_{\mathrm{C}}(E)\equiv\Lambda_{\mathrm{C}}(E)-i\pi\nu_{\mathrm{C}}(E)$, where
\begin{align}
\Lambda_{\mathrm{C}}(E)&=\tfrac{E}{2D^2}\ln\Bigl|\tfrac{E^2(E^2-\gamma_1^2)}{(D^2-E^2)^2}\Bigr|+\tfrac{\gamma_1\Delta_\mathrm{C}}{2D^2}\ln\Bigl|\tfrac{E+\gamma_1}{E-\gamma_1}\Bigr|\,,\\
\nu_{\mathrm{C}}(E)&=\sum\limits_{{\mu=\pm}}\tfrac{|E|-\mu\Delta_\mathrm{C}\gamma_1}{2D^2}\,\Theta\bigl(D-|E|\bigr)\Theta\bigl(|E|-\mu \gamma_1\bigr)\,.\label{eq:projected DOS}
\end{align}
The symbol $D=\sqrt{\sqrt{3}\pi}\gamma_0\simeq 6\,$eV denotes the effective bandwidth, and $\Delta_\mathrm{C}=0$ for $\mathrm{C_d}$-site and $\Delta_\mathrm{C}=1$ for $\mathrm{C_{nd}}$-site in the BLG case, respectively. By setting $\gamma_1 = 0$ the above formulas apply to the SLG case\cite{Kochan2014:PRL}.

The adsorption of fluorine on graphene induces resonance levels in the graphene spectrum which directly affect the relaxation rates. We determine the resonance energy, i.e. the energy at which an incoming electron resonantly scatters off the impurity, from the perturbed DOS per atom and spin which is given by
\begin{align}\label{eq:supp_pdos}
\varrho_{\mathrm{C}}(E)=\sum_{\mu=\pm}\varrho_0^\mu(E)-{(\eta/\pi)}\,\frac{1}{4}\mathrm{Im}\sum_\ell&\bigl\{\bigl[-\tfrac{d}{dE}G_{\mathrm{C}}(E)\bigr]\times\bigr.\nonumber\\
&\bigl.\times\left(2\ell+1\right)\mathrm{T}_\ell(E)\bigr\}\,,
\end{align}
where, $\varrho_0^\mu(E)=(2|E|-\mu \gamma_1)/(4D^2)\,\Theta\bigl(D-|E|\bigr)\,\Theta\bigl(|E|-\mu \gamma_1\bigr)$ is the unperturbed BLG DOS per atom and spin for the high ($\mu=+$) and low ($\mu=-$) energy band, respectively.

Using the T-matrix and the generalized Fermi golden rule, the spin-dependent relaxation rate at a given energy for given adatom concentration $\eta$ is obtained from\cite{Kochan2014:PRL,Kochan2015:PRL}
\begin{align}\label{eq:supp_model_relaxation}
\hspace{-1mm}\frac{1}{\tau_{\sigma\sigma'}^{\mathrm{C}}}=\frac{\eta}{2}\,\frac{2\pi}{\hbar}&\left\{\delta_{\sigma\sigma'}|T_1(E)|^2+\frac{1}{4}|T_1(E)+(\sigma\cdot\sigma')T_0(E)|^2\right\}\nonumber\\
&\times\frac{\left[\mathrm{P}_{\mathrm{C}}^{+}(E)\varrho_0^{+}(E)+\mathrm{P}_{\mathrm{C}}^{-}(E)\varrho_0^{-}(E)\right]^2}{\varrho_0^{+}(E)+\varrho_0^{-}(E)}\,.
\end{align}
Here, we introduced the projection factor $\mathrm{P}_\mathrm{C}^\mu(E)=2(|E|-\mu\Delta_\mathrm{C}\gamma_1)/(2|E|-\mu \gamma_1)\,\Theta\bigl(D-|E|\bigr)\,\Theta(|E|-\mu \gamma_1)$ which specifies the contribution of the site C to the low and high energy bands $\mu$ at a given energy $E$. In the SLG case with $\gamma_1=0$, one has correspondingly $\varrho_0^{+}(E)=\varrho_0^{-}(E)$ and $\mathrm{P}_\mathrm{C}^{+}(E)=\mathrm{P}_\mathrm{C}^{-}(E)$. The adatom concentration $\eta$ is defined as the number of adatoms divided by the number of carbon atoms in the structure. The quantity $\eta$ is related to the areal impurity concentration, $n_{\rm F}$, via $\eta^{\rm SL}=n_{\rm F} A_{\rm uc}/2$ for SLG and $\eta^{\rm BL}=n_{\rm F} A_{\rm uc}/4$ for BLG, where $A_{\rm uc}=3(\sqrt{3}/2)a_{cc}^2$ is the area of one graphene unit cell with the carbon-carbon distance $a_{cc}$.

From Eq.~(\ref{eq:supp_model_relaxation}) we obtain both the spin relaxation rate, $1/\tau^\mathrm{C}_s = 1/\tau^\mathrm{C}_{\uparrow\downarrow} + 1/\tau^\mathrm{C}_{\downarrow\uparrow}$, and the momentum relaxation rate, $1/\tau^\mathrm{C}_m = 1/\tau^\mathrm{C}_{\uparrow\uparrow} + 1/\tau^\mathrm{C}_{\uparrow\downarrow}$.
In the case of fluorinated BLG we assume that both dimer and nondimer sites contribute statistically equally to the relaxation and, therefore, the final spin relaxation rate is given by their unbiased average:
\begin{equation}\label{eq:supp_spin_rel}
1/\tau_{s(m)}\equiv 1/\bigl(2\tau_{s(m)}^{\mathrm{C_d}}\bigr)+1/\bigl(2\tau_{s(m)}^{\mathrm{C_{nd}}}\bigr)\,.
\end{equation}
We checked that the results presented in the main text of the paper do not change qualitatively under variation of the ratio of dimer and nondimer adsorption positions. Finally, the effect of charge puddles present in the experimental samples are taken into account by a Gaussian broadening of the relaxation rates by $\Sigma_{\rm eh}$.

\paragraph{Charged impurity scattering:\label{par:chimp}}
For calculating the momentum relaxation rate for charged fluorine scattering, we employ the model of Refs.~\onlinecite{Hwang2007,Hwang:PRB2008,dasSarma2010,dasSarma2011} in the approximation of zero temperature. For simplicity, we further assume that each fluorine adatom carries a charge of $-e$, neglect the finite distance of fluorine to graphene~\cite{Irmer2015:PRB}, and set the relative permittivity of the fluorine environment to graphene on SiO$_2$~\cite{Adam2007}. Both the short (resonant scattering) and long range (charged impurity) contributions to the momentum relaxation rate are then combined by the Matthiesen's rule to obtain $\tau_{\rm m, eff}^{-1}$ in Fig.~\ref{fig:rateRatio_bl}(b).

\begin{acknowledgments}
Work at Regensburg is supported by DFG SFB 689, SFB 1277 (A09), and the European Unions Horizon 2020 research and
innovation program under Grant No. 785219. Work at Penn State is supported by NSF (grant nos. DMR-1708972 and DMR-1506212). M. Gmitra acknowledges support by MSVVaS SR 90/CVTISR/2018 and
VVGS-2018-887. We thank Aires Ferreira for helpful discussions.
\end{acknowledgments}

%-------------------------------------------------------------------------------------

\bibliography{paper}

%-------------------------------------------------------------------------------------

\end{document}